\documentclass[rapids]{jfm}
\usepackage[resetlabels]{multibib}

\usepackage[usenames, dvipsnames]{color}
\usepackage{ar}
\usepackage{amsmath,color,graphicx}%
\usepackage{amsbsy}
\usepackage{latexsym}
\usepackage[normalem]{ulem}
\usepackage[english]{babel}
\usepackage{ifsym}
\usepackage{bbding}
\usepackage{wasysym}
\usepackage{epsfig} 
\usepackage{epsf,psfig}
\usepackage{epstopdf}
\usepackage{multirow}
\usepackage{float}
\usepackage[english]{babel}

\usepackage{verbatim}
\usepackage{soul}
\usepackage[toc,page]{appendix}
\usepackage{setspace}
\usepackage{threeparttable}
\usepackage{booktabs,array}
\usepackage{mathtools}
\usepackage{siunitx}
\usepackage{setspace}
\usepackage[numbers]{natbib}
\usepackage{authblk}

\definecolor{cobalt}{rgb}{0.0, 0.28, 0.67} %
\definecolor{cadetblue}{rgb}{0.37, 0.62, 0.63} %
\definecolor{carnelian}{rgb}{0.7, 0.11, 0.11}
\definecolor{cadmiumgreen}{rgb}{0.0, 0.42, 0.24}
\definecolor{orange}{rgb}{1,0.5,0}
\definecolor{purple}{rgb}{0.57,0,0.86}

\def\myPathTextAbove#1#2#3#4{
\draw [-LaTeX,thick,postaction={decorate,
               decoration={
                         raise=1ex,
                         text along path,
                         text align={center},
                         text={%
                                     |\color{black}| {#1} }
                              }
                        }
	  ] #2 to [bend left=#4] #3;
}

\usepackage{tikz,pgfplots,pstool}
\usepackage{pgfkeys} 
\pgfplotsset{compat=1.13}

\usetikzlibrary{calc, 
	decorations.markings, 
	decorations.text,
	decorations.pathreplacing,
	shapes.multipart,
	positioning,
	arrows.meta,
	arrows}

\usepackage[colorinlistoftodos,prependcaption,textsize=tiny]{todonotes}

\usetikzlibrary{fadings}

\tikzfading[
  name=fade out,
  inner color=transparent!0,
  outer color=transparent!100
]
 
 \makeatletter
 \graphicspath{{/figures/}}
 \makeatother
 
 \usetikzlibrary{external}
	  \tikzset
  {
    myCircleL/.style=
    {
      red,
     path fading=fade out,
    }
  }
  
	  \tikzset
  {
    myCircleR/.style=
    {
      blue,
      path fading=fade out,
    }
  }  

\def\myVortexL#1#2#3#4{
	\begin{scope}[shift = {#1}, scale=0.1]
					\fill[myCircleL] (0,0) circle (3*#2);
		\draw [fill=black] (0,0) circle [radius=0.4*#2] node [below=.5em] {#4};
		\draw[->] ([shift=(0:#2)] 0,0) arc (0:270:#2);
	\end{scope}
}

\def\myVortexR#1#2#3#4{
	\begin{scope}[shift = {#1}, scale=0.1]
			\fill[myCircleR] (0,0) circle (3*#2);
		\draw [fill=black] (0,0) circle [radius=0.4*#2] node [below=.5em] {#4};
		\draw[->] ([shift=(0:#2)] 0,0) arc (0:-270:#2);
	\end{scope}
}

\usepackage{bm}
\def\e{\textrm{e}}
\def\d{\textrm{d}}
\def\i{\textrm{i}}

\usepackage{physics}
\usepackage{xcolor,colortbl}

\definecolor{themecolour}{rgb}{0.64, 0.76, 0.68}
\definecolor{mygray}{rgb}{0.7, 0.7, 0.7}
\definecolor{tablegray}{rgb}{0.75, 0.75, 0.75}
\definecolor{confcol}{RGB}{108, 172, 228}
\definecolor{camblue}{RGB}{108, 172, 228}
\definecolor{camred}{RGB}{213, 0, 50}
\definecolor{camnavy}{RGB}{0, 60, 113}

\definecolor{myblue}{rgb}{0 ,  0.4470 , 0.7410}
\definecolor{myorange}{rgb}{0.8500,    0.3250,    0.0980}
\definecolor{myyellow}{rgb}{0.9290,    0.6940,    0.1250}
\definecolor{mypurple}{rgb}{ 0.4940,    0.1840,    0.5560}
\definecolor{myred}{rgb}{     0.6350 ,   0.0780 ,   0.1840}
\definecolor{mygreen}{rgb}{         0.4660  ,  0.6740   , 0.1880}

\definecolor{prsared}{RGB}{ 219,25,73}

\definecolor{mygrey}{rgb}{ .9,.9,.9}

\definecolor{confblue}{rgb}{0.75, 0.72, 0.95}
\definecolor{confgreen}{rgb}{0.76, 1, 0.74}

\newlength{\xshift}
\newlength{\yshift}
\newlength{\fwidth}
\usepackage{lipsum}

\usepackage{url}

\newlength{\myheight}
\newlength{\mywidth}
\newlength{\myww}
\newlength{\myhh}

\newsavebox{\imagebox}
\newsavebox{\fadedimagebox}

\newcommand{\niceFigure}[4]{

	\savebox{\imagebox}{\begin{tikzpicture}
		\node[inner sep = 0pt] at (0, 0) {\includegraphics[width = .6\linewidth]{#4}};
		\end{tikzpicture}}%
	\settowidth{\myww}{\usebox{\imagebox}}
	\settoheight{\myhh}{\usebox{\imagebox}}
	\savebox{\fadedimagebox}{\begin{tikzpicture}
		\node[inner sep = 0pt] at (0, 0) {\includegraphics[width = \myww]{#4}};
		\end{tikzpicture}}%

	\begin{subfigure}[t]{.25\linewidth}
		\centering\raisebox{\dimexpr.5\ht\fadedimagebox-.5\height}{%
			\includegraphics[width = \linewidth,height = \linewidth]{#3}} %
		\caption{$\zeta$-plane}
	\end{subfigure}\quad \begin{tikzpicture}[remember picture] \myPathTextAbove{$z = f(\zeta)$}{(-.5,0)}{(.5,0)}{20}; \node at (0,-2) {};
	\end{tikzpicture} \quad
	\begin{subfigure}[t]{.6\linewidth}
		\centering\usebox{\fadedimagebox}%
		\caption{$z$-plane}
	\end{subfigure}
}
\shorttitle{Exact solutions for ground effect} %
\shortauthor{Baddoo, Kurt, Ayton \& Moored} %

\title{Exact solutions for ground effect }

\author
 {
 Peter J. Baddoo\aff{1}
  \corresp{\email{baddoo@damtp.cam.ac.uk}},
  Melike Kurt\aff{2}, Lorna J. Ayton\aff{1}, 
  \& 
  Keith W. Moored\aff{2}
  }

\affiliation
{\aff{1}
Department of Applied Mathematics and Theoretical Physics, University of Cambridge, Cambridge, UK
\aff{2}
Department of Mechanical Engineering, Lehigh University, Bethlehem, PA 18015, USA

}
\usepackage{subcaption}
\newcites{Supp}{Supplementary material references}

\begin{document}
\maketitle
\begin{abstract}
``Ground effect'' refers to the enhanced performance enjoyed by fliers or swimmers operating close to the ground. We derive a number of exact solutions for this phenomenon, thereby elucidating the underlying physical mechanisms involved in ground effect. Unlike previous analytic studies, our solutions are not restricted to particular parameter regimes such as ``weak'' or ``extreme'' ground effect, and do not even require thin aerofoil theory. Moreover, the solutions are valid for a hitherto intractable range of flow phenomena including point vortices, uniform and straining flows, unsteady motions of the wing, and the Kutta condition. We model the ground effect as the potential flow past a wing inclined above a flat wall. The solution of the model requires two steps: firstly, a coordinate transformation between the physical domain and a concentric annulus, and secondly, the solution of the potential flow problem inside the annulus. We show that both steps can be solved by introducing a new special function which is straightforward to compute. Moreover, the ensuing solutions are simple to express and offer new insight into the mathematical structure of ground effect. In order to identify the missing physics in our potential flow model, we compare our solutions against new experimental data. The experiments show that boundary layer separation on the wing and wall occurs at small angles of attack, and we suggest ways in which our model could be extended to account for these effects.
\end{abstract}


\section{Introduction} \label{Sec:introduction}

Understanding ground effect is of intense interest to researchers seeking to comprehend natural flight \citep{rayner1991aerodynamics} and design efficient vehicles \citep{rozhdestvensky2006wing}. The performance improvements due to ground effect are well documented, and include enhancements in the lift-to-drag ratio \citep{Ahmed2002} and propulsive efficiency \citep{quinn2014unsteady}. Despite the appeal of understanding ground effect, associated analytic studies have been limited; some progress has been made through asymptotic analyses in regimes such as extreme ground effect \citep{Widnall1970, tuck1980nonlinear}, weak ground effect \citep{iosilevskii2008asymptotic}, and linearised thin aerofoils \citep{Katz2001,Tomotika1933}, but a uniformly valid theory is still lacking. In particular, there are no analytic studies of moderate ground effect, despite experimental data indicating that this regime exhibits unique physical behaviour such as flow-mediated equilibria \citep{Kurt2019}. Moreover, previous analytic works have only considered simple physical flow phenomena such as background uniform flows, linearised wakes, or small amplitude motions of the wing.

Perhaps one reason for the lack of analytic studies is the complicated topology associated with ground effect: the domain is \emph{doubly connected} as one must model both the wing and the ground. Although such domains have historically proved resilient to analytic treatments, recent work \citep{Crowdy2006,Crowdy2010,Crowdy2020} has elucidated the underlying mathematical structure of more general multiply connected domains. In particular, \cite{Crowdy2010} demonstrated the relevance of a special transcendental function -- known as the ``\emph{Schottky--Klein prime function}'' -- for solving fluid problems in multiply connected domains. By adapting the approach of \cite{Crowdy2010}, we account for the doubly connected topology of ground effect by using a special case of the prime function, thereby allowing exact solutions for the  potential flow problem. Accordingly, our theory captures all asymptotic regimes and is not limited to a linearised geometry. Additionally, our solution is fully capable of modelling all potential flow phenomena, such as point vortices to represent a shedding wake.

Our mathematical solution involves two steps. The first step is to determine an appropriate coordinate transformation (i.e. a conformal map) from a doubly connected annulus to the physical domain of interest. We offer several suggestions for such mappings, including a new analogue of the classical Joukowski mapping \citep{Joukowsky1910} for ground effect. The second step in the solution is the construction of the complex potential function inside the annular domain, which we perform using the ``new calculus of vortex dynamics'' by \cite{Crowdy2010}. We provide closed form representations of the solution for each step.

The remainder of the paper is arranged as follows. In section \ref{Sec:model} we present our mathematical model for ground effect. In section \ref{Sec:solution}, we present the mathematical solution to the model. We begin by presenting analytic expressions for some conformal maps from an annulus to a target physical domain in section \ref{Sec:conformal}. We then derive solutions for a range of physically relevant flows inside the annulus in section \ref{Sec:library}. In section \ref{Sec:examples},  we compare the potential flow solution to experimental results through interrogating the circulation for a range of distances and angles of attack. Finally, in section \ref{Sec:conclusion} we summarise the paper and suggest directions of future research.

\section{Mathematical model} \label{Sec:model} \label{Sec:ComPot}
We consider a two-dimensional, inviscid and incompressible flow with fluid velocity $\boldsymbol{u}$ in a physical $z$-plane where $z = x + \i y$. We additionally assume that the flow is irrotational except at discrete points where there may be point vortices. Therefore, we may define a  complex potential function
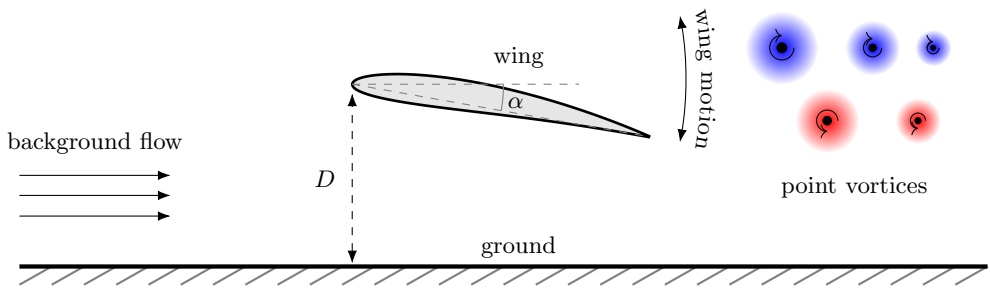
\begin{figure}
	\centering
	\begin{tikzpicture}[scale=.8]

	\def\aerofoilLength{5}
	\def\aerofoilHeight{3}
	\coordinate (1a) at (-\aerofoilLength/2,\aerofoilHeight);
	
	\def\numLines{30}
	\node at (0,5) {};
	\node at (0,-1) {};
	\foreach \x in {1,...,\numLines}
	\draw[thick,gray] ($(-10+16*\x/ \numLines-16/\numLines,-.3)$)--($(-10+16*\x/\numLines, 0)$);
	\draw[ultra thick] (-10,0)--(6,0);
	
\begin{scope}[shift = {(-2,0)}]	

\draw[thick, rotate around={-10:(-\aerofoilLength/2,\aerofoilHeight)},shift = {(-\aerofoilLength/2,\aerofoilHeight)},scale = \aerofoilLength,fill=mygrey,line width = 1pt] plot file{figures/aero3.dat};

\draw[dashed, gray, rotate around={-10:(-\aerofoilLength/2,\aerofoilHeight)},shift = {(-\aerofoilLength/2,\aerofoilHeight)},scale = \aerofoilLength,dashed] (0,0)--(1,0);

	\draw[dashed, gray] (-\aerofoilLength/2,\aerofoilHeight)--(.5*\aerofoilLength/2,\aerofoilHeight);

	\draw[dashed, Latex-Latex] (-\aerofoilLength/2,.05)--(-\aerofoilLength/2,\aerofoilHeight*0.95) node[midway, left=3pt] {$D$};

	\node at (-\aerofoilLength/2+2.7,\aerofoilHeight-.3) {$\alpha$};
   \draw [gray,domain=-10:0, shift = {(-\aerofoilLength/2,\aerofoilHeight)}] plot ({2.5*cos(\x)}, {2.5*sin(\x)});
\node[above = 3pt] at (.25,\aerofoilHeight) {\footnotesize wing};
\node[above] at (.25,0) {ground};
	\coordinate (bmeVortex) at (\aerofoilLength,{1.8*\aerofoilHeight});
\draw[Latex-Latex, postaction={decorate, decoration={text along path, raise=6pt, text align={align=center}, text={wing motion}, reverse path}}] ($({-\aerofoilLength/2+1.1*\aerofoilLength*cos(10)},{-1.1*\aerofoilLength*sin(10)+\aerofoilHeight})$) arc (-10:10:1.2*\aerofoilLength);
\end{scope}

\begin{scope}[shift = {(\aerofoilLength/2+.1,0)}]
	\myVortexR{(0,\aerofoilHeight+.6)}{2}{1}{}
		\myVortexL{(.75,\aerofoilHeight-.6)}{1.75}{1}{}
	\myVortexR{(1.5,\aerofoilHeight+.6)}{1.5}{1}{}
	\myVortexL{(2.25,\aerofoilHeight-.6)}{1.25}{1}{}
		\myVortexR{(2.5,\aerofoilHeight+.6)}{1}{1}{}
	\node at (1.2,1.3) {\footnotesize point vortices};
	\end{scope}
	
		\draw[-Latex] (-10,1.5)--(-7.5,1.5) node[midway, above=5pt] {background flow};
	\draw[-Latex] (-10,1.17)--(-7.5,1.17);
    \draw[-Latex] (-10,.83)--(-7.5,.83);
	\end{tikzpicture} \vspace{-.5cm}
	\caption{The mathematical model in the present work. An wing is in ground effect in a background flow. The chord length is non-dimensionalised to unity, the angle of attack is {$\alpha$}, and the distance between the leading-edge and the ground is {$D$}. The origin is placed on the ground in-line with the leading-edge. A representative pitching motion with the associated von K\'arm\'an vortex street is illustrated. The \textcolor{red}{\textbf red} spirals correspond to vortices with positive circulation whereas \textcolor{blue}{\textbf blue} spirals denotes vortices with negative circulation.} \label{Fig:Model}
\end{figure}
\begin{align*}
w(z) = \phi(z) + \i \psi(z),
\end{align*}
where $\phi$ and $\psi$ are the velocity potential and streamfunction in the physical $z$-plane, respectively. The complex potential function is harmonic so that $\nabla^2 w=0$.

We now outline the geometry of our model, which is illustrated in figure \ref{Fig:Model}. We model our flier or swimmer as a wing of non-dimensional length $1$. The wing is inclined at angle of attack $\alpha$ and the distance of the leading edge to the ground is $D$. The wing may, in principle, take any shape: by using conformal mappings, our solution is valid for any wing profile. We present closed-form mapping functions for several wing shapes of practical interest in section \ref{Sec:conformal}. We allow the wing to undergo pitching and heaving motions so that $\alpha$ and $D$ may depend on time. We also consider a background flow, and may include point vortices in the flow to represent discretised vorticity. The ground and the wing are impermeable and thus the flow must satisfy a no-flux condition on each of these boundaries. Length and velocity scales have been non-dimensionalised with respect to the chord length and upstream velocity respectively.

\section{Mathematical solution} \label{Sec:solution}
As mentioned in the introduction, a significant obstacle associated with the current mathematical model is the fact that it is \emph{doubly connected}. This topology presents a challenge insofar as the mathematical solution is concerned since most analytic methods are only applicable to simply connected domains. To overcome this difficulty we introduce a special function known as the ``$P$-function'' which may be simply expressed as the infinite product
	\begin{align}
	P(\zeta)  &= (1 - \zeta) \prod_{k=1}^{\infty} \left(1 - q^{2k} \zeta \right) \cdot \left( 1- q^{2k} \zeta^{-1}\right), \label{Eq:SKprod}
	\end{align}
for $0<q<1$. We suppress the dependence on $q$ for notational convenience. The \mbox{$P$-function} is a special case of the Schottky--Klein prime function, which has found relevance as a fundamental tool for solving fluid problems in multiply connected domains \citep{Crowdy2010,Crowdy2020}. In particular, the significance of the \mbox{$P$-function} is that it is effectively a generalisation of the function $(1 - z)$ to doubly connected domains \citep{Baker1897}. A full discussion of this fact is well beyond the scope of the current article, but for now it is sufficient to note that the \mbox{$P$-function} is analytic in the annulus $q< |\zeta|< q^{-1}$, inside which it never vanishes except for a simple zero at $\zeta = 1$. Further details of the $P$-function are provided in section \ref{Sec:Pfun} of the supplementary material.

Remarkably, both the conformal maps \emph{and} the complex potentials for a range of flows can be expressed exclusively in terms of the $P$-function. %

\subsection{Conformal maps} \label{Sec:conformal}
\begin{figure}
	\centering
		\begin{tikzpicture}[scale=1]
	
	\begin{scope}[scale = .6]
	\path [path fading = circle with fuzzy edge 15 percent, fill=mygrey, even odd rule] (0,0) circle (3) (0,0) circle (4);
	\coordinate (v) at (1.5,1.5);
	
	\draw [black,fill=white] (0,0) circle (3);
	\draw [fill=mygrey] (0,0) circle [radius=1.3];
	
	\draw [ thick, red, line width = 2pt ] (0,0) circle (1.3);
	\draw [ thick, blue, line width = 2pt ] (0,0) circle (3);

	\draw[Latex-Latex, dashed] (0,0)--(3*0.9659,3*0.2588) node[above,midway] {$1$};
	\node[right] at (3*0.9659,3*0.2588) {$C_0$};
	\draw[Latex-Latex, dashed] (0,0)--(0,-1.3) node[left,midway] {$q$};
	\node[left] at (-1.3,0) {$C_1$};
	\end{scope}

	\begin{scope}[shift = {(4.5,1.25)},scale =.6]
	\coordinate (bl) at (-3,-2);
		\coordinate (br) at (3,-2);
	\coordinate (blg) at (-3,-1);
	\coordinate (brg) at (3,-1);
	\coordinate (tlg) at (-3,1);
	\coordinate (trg) at (3,3);
	\coordinate (tlb) at (3,-1.5);
	\coordinate (trb) at (-3,1.5);
	
	\coordinate (v1) at (0,.5);
	
	\path [fill=mygrey, path fading = south] (blg) rectangle (tlb);
	
	\draw [blue,  ultra thick] (blg)--(brg);
	\draw [red,  line width = 2pt] (-1,1)--(1,0.5);
		\path [fill=white,path fading=east] ($(bl)-(.01,0)$) rectangle ($(blg)+(1,1)$);
		\path [fill=white,path fading=west] ($(br)+(0.01,0)$) rectangle ($(brg)+(-1,1)$);

\node at (0,-2) {flat plate};
	\end{scope}
	
	\begin{scope}[shift = {(-4.5,1.25)},scale =.6]
	\coordinate (bl) at (-3,-2);
		\coordinate (br) at (3,-2);
	\coordinate (blg) at (-3,-1);
	\coordinate (brg) at (3,-1);
	\coordinate (tlg) at (-3,1);
	\coordinate (trg) at (3,3);
	\coordinate (tlb) at (3,-1.5);
	\coordinate (trb) at (-3,1.5);
	
	\coordinate (v1) at (0,.5);
	
	\path [fill=mygrey, path fading = south] (blg) rectangle (tlb);
	
	\draw [blue,  ultra thick] (blg)--(brg);
	\draw [red, fill = mygrey, line width = 2pt] (0,.5) circle (1);
		\path [fill=white,path fading=east] ($(bl)-(.01,0)$) rectangle ($(blg)+(1,1)$);
		\path [fill=white,path fading=west] ($(br)+(0.01,0)$) rectangle ($(brg)+(-1,1)$);

	\node at (0,-2) {\footnotesize circular wing};

	\end{scope}

	\begin{scope}[shift = {(-4.5,-1.25)},scale =.6]
	\coordinate (bl) at (-3,-2);
		\coordinate (br) at (3,-2);
	\coordinate (blg) at (-3,-1);
	\coordinate (brg) at (3,-1);
	\coordinate (tlg) at (-3,1);
	\coordinate (trg) at (3,3);
	\coordinate (tlb) at (3,-1.5);
	\coordinate (trb) at (-3,1.5);
	
	\coordinate (v1) at (0,.5);
	
	\path [fill=mygrey, path fading = south] (blg) rectangle (tlb);
	
	\draw [blue,  ultra thick] (blg)--(brg);
	\draw [gray,dashed,line width = .5pt,domain=0:360, shift = {(0,.5)}] plot ({1*cos(\x)}, {1*sin(\x)});
	   \draw [red,line width = 2pt,domain=-150:00, shift = {(0,.5)}] plot ({cos(\x)}, {sin(\x)});
		\path [fill=white,path fading=east] ($(bl)-(.01,0)$) rectangle ($(blg)+(1,1)$);
		\path [fill=white,path fading=west] ($(br)+(0.01,0)$) rectangle ($(brg)+(-1,1)$);

	\node at (0,-2) {\footnotesize circular arc wing};

	\end{scope}

	\begin{scope}[shift = {(4.5,-1.25)},scale =.6]
	\coordinate (bl) at (-3,-2);
		\coordinate (br) at (3,-2);
	\coordinate (blg) at (-3,-1);
	\coordinate (brg) at (3,-1);
	\coordinate (tlg) at (-3,1);
	\coordinate (trg) at (3,3);
	\coordinate (tlb) at (3,-1.5);
	\coordinate (trb) at (-3,1.5);
	
	\coordinate (v1) at (0,.5);
	
	\path [fill=mygrey, path fading = south] (blg) rectangle (tlb);
	
	\draw [blue,  ultra thick] (blg)--(brg);
		   \draw [gray,dashed,line width = .5pt,domain=0:180, shift = {(0,-1)}] plot ({1.5*cos(\x)}, {1.5*sin(\x)});
	   \draw [red,line width = 2pt,domain=30:120, shift = {(0,-1)}] plot ({1.5*cos(\x)}, {1.5*sin(\x)});

		\path [fill=white,path fading=east] ($(bl)-(.01,0)$) rectangle ($(blg)+(1,1)$);
		\path [fill=white,path fading=west] ($(br)+(0.01,0)$) rectangle ($(brg)+(-1,1)$);

	\node at (0,-2) {\footnotesize centered circular arc wing};

	\end{scope}
	
	\draw [-Latex] (1.75,1.75) to [out=45,in=155] (3,2);
			\draw [-Latex] (-1.75,1.75) to [out=135,in=25] (-3,2);
		\draw [ -Latex] ($({1.75*cos(205)/cos(45)},{1.75*sin(205)/cos(45)})$) to [out=205,in=0] ($({2.2*cos(202)/cos(45)},{2.2*sin(202)/cos(45)})$);
		\draw [ -Latex] ($({1.75*cos(-25)/cos(45)},{1.75*sin(-25)/cos(45)})$) to [out=-25,in=180] ($({2.2*cos(-22)/cos(45)},{2.2*sin(-22)/cos(45)})$);
	\end{tikzpicture}
	\caption{The four conformal mappings for ground effect. Each map relates an  annular domain ($\zeta$-space) to the target physical domain ($z$-space). The interior circle $C_1$ (coloured \textcolor{red}{\textbf red}) is mapped to the wing whereas the unit circle $C_0$ (coloured \textcolor{blue}{\textbf blue}) is mapped to the ground plane. Areas outside the domain of definition are shaded in gray.} \label{Fig:ConfMap}
\end{figure}
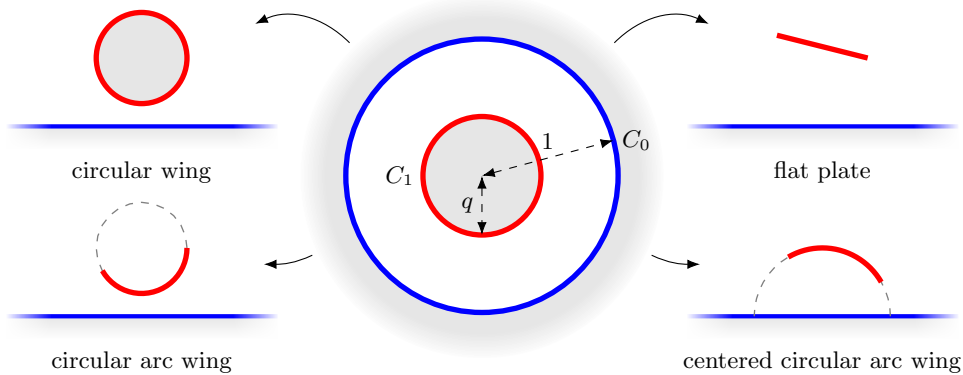

The complicated geometry of the ground effect problem means that it is expedient to construct an analytic solution in a circular domain (labelled $D_\zeta$) and then map the solution to the physical domain of interest (labelled $D_z$). Without loss of generality, we may define the circular domain to be a concentric annulus bounded by the unit disc. We denote the unit disc as $C_0$ whilst the interior circle is denoted by $C_1$ and has radius $q$ as illustrated in figure \ref{Fig:ConfMap}. We seek a mapping, $f$, that relates $C_0$ to the ground plane and $C_1$ to the wing so that we may write $z = f(\zeta)$.

We now present four conformal mappings of relevance to ground effect. With the exception of the first mapping \eqref{Eq:Mobius}, every mapping presented herein is novel especially in their application to ground effect. In the following formulae, the constant $A$ is used to rotate or rescale the domain, and the constant $s$ is used to shift the domain. Further examples of the mappings are illustrated in section \ref{Sec:confProof} of the supplementary material. 

\subsubsection{Circular wing map}
The first mapping is the standard M\"obius map
\begin{align}
    f(\zeta) & = \frac{1-q^2}{2\i q} \cdot \frac{\zeta+1}{\zeta-1} + s,
    \label{Eq:Mobius}
\end{align}
which maps $C_0$ to the real axis and $C_1$ to a circle of unit radius centered at $\i(q+q^{-1}))/2$.
\subsubsection{Flat plate map}
The second mapping takes the form
\begin{align}
f(\zeta) = A \frac{P(\zeta\e^{2 \i \alpha})}{P(\zeta)} +s, \label{Eq:ConfMap}
\end{align}
and maps $C_0$ to the real axis and $C_1$ to a flat plate of unit length oriented at angle of attack $\alpha$. The distance of the plate from the ground is determined by the radius of the interior circle $q$. This mapping -- which has not been seen before in the ground effect literature -- is particularly significant because it is a natural generalisation of the famous Joukowski map which relates the unit disc to a single slit.  Modelling aerofoils as flat plates is ubiquitous in classical \citep{Katz2001} and data-driven \citep{Darakananda2018a} aerodynamics, and the inclusion of the ground plane here is a major extension: the plethora of studies using the Joukowski map can now be adapted for ground effect.

In the special case where the plate is parallel to the ground, the mapping \eqref{Eq:ConfMap} degenerates to
\begin{align}
f(\zeta) =  A \zeta \frac{P^\prime (\zeta)}{P(\zeta)} +s. \label{Eq:confMap0}
\end{align}
\subsubsection{Circular arc wing map}
The third mapping transplants $C_0$ to the real axis and $C_1$ to a circular arc with center located in the upper half plane. The mapping takes the functional form
\begin{align}
f(\zeta) & = A \frac{P(\zeta \overline{\gamma})}{P(\zeta/\gamma)P(\overline{\gamma}) - P(1/\gamma)P(\zeta \overline{\gamma})}+s, \label{Eq:circArcMap}
\end{align}
for any $\gamma\in D_\zeta$. 
\subsubsection{Centered circular arc wing map}
The fourth mapping sends $C_0$ to the real line and $C_1$ to a circular arc centered at the origin. The mapping may be expressed as
\begin{align}
f(\zeta) & = A \frac{P(\sqrt{\zeta}\e^{\i \phi})P(-\sqrt{\zeta}\e^{\i \phi})P(q \sqrt{\zeta})P(-q \sqrt{\zeta})}{P(q \sqrt{\zeta} \e^{\i \phi}) P(- q\sqrt{\zeta} \e^{\i \phi})P(\sqrt{\zeta})P(-\sqrt{\zeta})}+s, 
\label{Eq:centCircMap}
\end{align}
for $\phi \in [0,\pi]$.

The ensuing potential flows solutions are invariant under conformal maps, and are therefore applicable to any geometry. In this paper we only present mappings for simple wing shapes since they admit tractable closed form representations; more realistic wing shapes will require numerical conformal mapping procedures.

Finally, it is important to note that each map has a simple pole at $\zeta=1$, i.e.
\begin{align}
    f(\zeta) \sim \frac{a_{\infty}}{\zeta - 1}, \qquad \qquad \textrm{as }\zeta \rightarrow 1,
\end{align}
for constant $a_{\infty}$ defined for each map in section \ref{Sec:residue} of the supplementary material.
\subsection{Complex potential} \label{Sec:library}
Since we have a conformal map from the annular domain to the physical domain, it is sufficient to calculate the complex potential in the annular domain and then map this solution to the physical domain. We express the complex potential in the $\zeta$-plane as $W(\zeta)=w(z(\zeta))$. To construct $W$ we use the new calculus of vortex dynamics proposed by \cite{Crowdy2010}, which provides a framework for the calculation of instantaneous complex potentials associated with more general multiply connected domains. We present the instantaneous solutions for five scenarios of interest: circulatory flow, point vortices, uniform flow, straining flow, and unsteady pitching and heaving motions of the wing. In all but the last case, the no-flux condition is equivalent to enforcing that the wing is a streamline and therefore the complex potential takes constant imaginary part on both $C_0$ and $C_1$. 

We present the solutions here with little explanation; further details may be found in \cite{Crowdy2010}.

 \subsubsection{Specifying circulation around the wing: {$W_\Gamma(\zeta)$}}

We begin with the simplest possible non-trivial flow. Suppose that the flow has no singularities and decays in the far field, but the wing has circulation $\Gamma$. We may write the corresponding complex potential as
\begin{align}
W_\Gamma( \zeta) = \frac{\Gamma}{2 \pi \i} \log \left(\zeta\right).
 \end{align}
 It is simple to verify that $W_{\Gamma}$ takes constant imaginary part on $C_0$ and $C_1$. Furthermore, we may note that the circulation around the ground and the wing must be equal and opposite when there are no singularities in the flow. Later, we will use $W_\Gamma$ to set the circulation around the wing to enforce the Kutta condition.

\subsubsection{Contribution from point vortices: {$W_{V}(\zeta)$}}
Suppose that there is now a point vortex of unit strength embedded in the flow at $\beta$ in the annular domain. The corresponding Green's function $G$ then satisfies
  \begin{align}    
  \nabla^2_{\zeta} G(\zeta,\beta) &= -\delta(\zeta - \beta), \label{Eq:green}
\end{align}
for $\zeta \in D_{\zeta}$, in addition to the no-flux boundary condition. Due to \cite{Crowdy2010}, we know that the solution of \eqref{Eq:green} is the \emph{hydrodynamic Green's function}, which may be expressed in terms of the $P$-function as
\begin{align}
G(\zeta, \beta) = \frac{1}{2 \pi \i} \log \left(\frac{P(\zeta/\beta)}{|\beta| P(\zeta \overline{\beta})}\right), \label{Eq:G0}
\end{align}
where the overbar denotes the complex conjugate. Moreover, \cite{Crowdy2010} showed that $G$ produces circulation~$-1$ around the object which is the image of $C_0$ (the ground) and zero circulation around $C_1$ (the wing). 

It is worth noting the structure of the Green's function in \eqref{Eq:G0} when compared to the Green's function for a wing in isolation. In the latter (simply connected) case, the circular domain is the unit disc and the corresponding Green's function may be constructed by placing an image vortex at the location of the physical vortex reflected in the disc. Conversely, in the doubly connected case \eqref{Eq:G0}, there is an infinite family of image vortices existing outside the annulus. For a fixed $\beta$, inspection of the infinite product formula \eqref{Eq:SKprod} for the $P$-function shows that the set $\{q^{2k}\beta  , \;\; q^{2k}/\overline{\beta} : k \in \mathbb{Z} \setminus \{0\} \}$ represents the full family of image vortices. Accordingly, there are two infinite families of vorticies in the physical domain: one located inside the wing and the other located under the ground.

The complex potential for $N$ vortices located at $\beta_j\in D_\zeta$ with strength $\kappa_j$ may simply be expressed as
\begin{align}
W_{V}( \zeta) &= \sum_{j = 1}^N \kappa_j G(\zeta,\beta_j).\label{Eq:W_infty}
\end{align}
An appropriate arrangement of vortices could be used to model the wake shed by the wing through adapting the analysis of \cite{Michelin2009}.
\subsubsection{Contribution from uniform flow: {$W_U(\zeta)$}}
We now consider the case where the flow is uniform in the far field. The complex potential for a uniform flow takes the form $w_U \sim z$ as $|z| \rightarrow \infty$ in the physical domain. In the annular domain, this condition becomes $W_U \sim a_{\infty}/(\zeta-1)$ as $\zeta \rightarrow 1$. Additionally, the imaginary part of the complex potential must take constant values on $C_0$ and $C_1$. By slightly adapting \cite{Crowdy2010}, we may write
\begin{align}
W_U(\zeta) &= \zeta \frac{a_{\infty}}{L} \cdot \frac{P^\prime(\zeta)}{P(\zeta)},
\end{align}
where $L = \prod_{k=1}^\infty (1-q^{2k})^2$ is a constant that ensures that the speed is unity at infinity.

\subsubsection{Contribution from straining flow: $W_S(\zeta)$}
By using a similar approach we may calculate the complex potential for a straining flow. In this case, the complex potential has the asymptotic behaviour $w_S \sim z^2$ as $|z| \rightarrow \infty$ in the physical plane, in addition to the usual no-flux boundary conditions. The complex potential in the $\zeta$-plane is given by
\begin{align*}
    W_S(\zeta) & = -\frac{a_{\infty}}{2 L^2} \left(\frac{\zeta P^\prime(\zeta)}{P(\zeta)} + \frac{\zeta^2  P^{\prime \prime}(\zeta)}{P(\zeta)}-2\left(\frac{\zeta P^\prime(\zeta)}{P(\zeta)} \right)^2 \right).
\end{align*}
Analogous results for higher-order flows (i.e. $w \sim z^n$ as $|z| \rightarrow \infty$) may be derived by taking parametric derivatives of the hydrodynamic Green's function \eqref{Eq:green}.
 
\subsubsection{Contribution from wing motion: {$W_M(\zeta)$}} 
We may also consider the case where the wing executes rigid body motions of arbitrary amplitude. We show in section \ref{Sec:kinematic-sup} of the supplementary material that the corresponding instantaneous complex potential evaluated at $\zeta \in C_1$ must now satisfy
    \begin{align}
    \Im\left[ W_M(\zeta) \right] &=  \Im \left[ \dot{\bar{D}}(z(\zeta)-D) + \i\frac{\dot{\alpha}}{2} \left| z(\zeta)-D\right|^2 \right] + I\equiv M(\zeta,\bar{\zeta}),
    \label{Eq:boundary}
    \end{align}
where the dot corresponds to derivatives with respect to time and $I$ is a constant defined in \eqref{Eq:Idef}. Additionally, the imaginary part of $W_M$ should vanish on $C_0$. Therefore, $W_M$ is an analytic function inside the annulus with prescribed imaginary part on the boundaries given by \eqref{Eq:boundary}. Finding such a function is known as the \emph{Schwarz problem}, and the solution is given by the Villat formula \citep{Crowdy2008}. Accordingly, the complex potential may be expressed in terms of the $P$-function as
\begin{align*}
 W_M (\zeta) &= -\frac{1}{\pi \i } \oint_{C_1} M(\zeta_1,\overline{\zeta}_1)   \cdot \frac{P^\prime(\zeta/\zeta_1)}{P(\zeta/\zeta_1)} \cdot \frac{\d \zeta_1}{\zeta_1^2}.
 \end{align*}
\subsubsection{Total complex potential}
In summary, the total complex potential for a wing in ground effect is simply given by combining the contributions from the various flow phenomena:
\begin{align*}
W &= W_{\Gamma}(\zeta)+W_{V}(\zeta) + W_U(\zeta) + W_S(\zeta) + W_M(\zeta).
\end{align*}
In its current form, the solution is not unique: thus far we have said nothing about how to fix the circulation around the wing. The Kutta condition is the usual strategy for determining the circulation in aerodynamic applications \citep{Eldredge2019a}. To enforce the Kutta condition, we restrict the velocity at the trailing edge to be finite by setting 
\begin{align}
\Gamma &= -{2 \pi \i \zeta_t}\left(\frac{\d W_{V}}{\d \zeta} (\zeta_t) + \frac{\d W_{U}}{\d \zeta} (\zeta_t) +\frac{\d W_{S}}{\d \zeta} (\zeta_t) +\frac{\d W_{B}}{\d \zeta} (\zeta_t) \right), \label{Eq:circ}
\end{align}
where $\zeta_t$ is the preimage of the trailing edge in the $\zeta$-plane.
\section{Experimental validation} \label{Sec:examples}
In this section we compare our exact ideal solution to experimental data that was collected from flow-visualization experiments. We consider the case of an inclined flat plate in ground effect embedded in a uniform flow. For the potential flow solution, the relevant conformal mapping is \eqref{Eq:ConfMap} and the circulation is given by \eqref{Eq:circ}.

\begin{figure}
\centering
\includegraphics[width=1\linewidth]{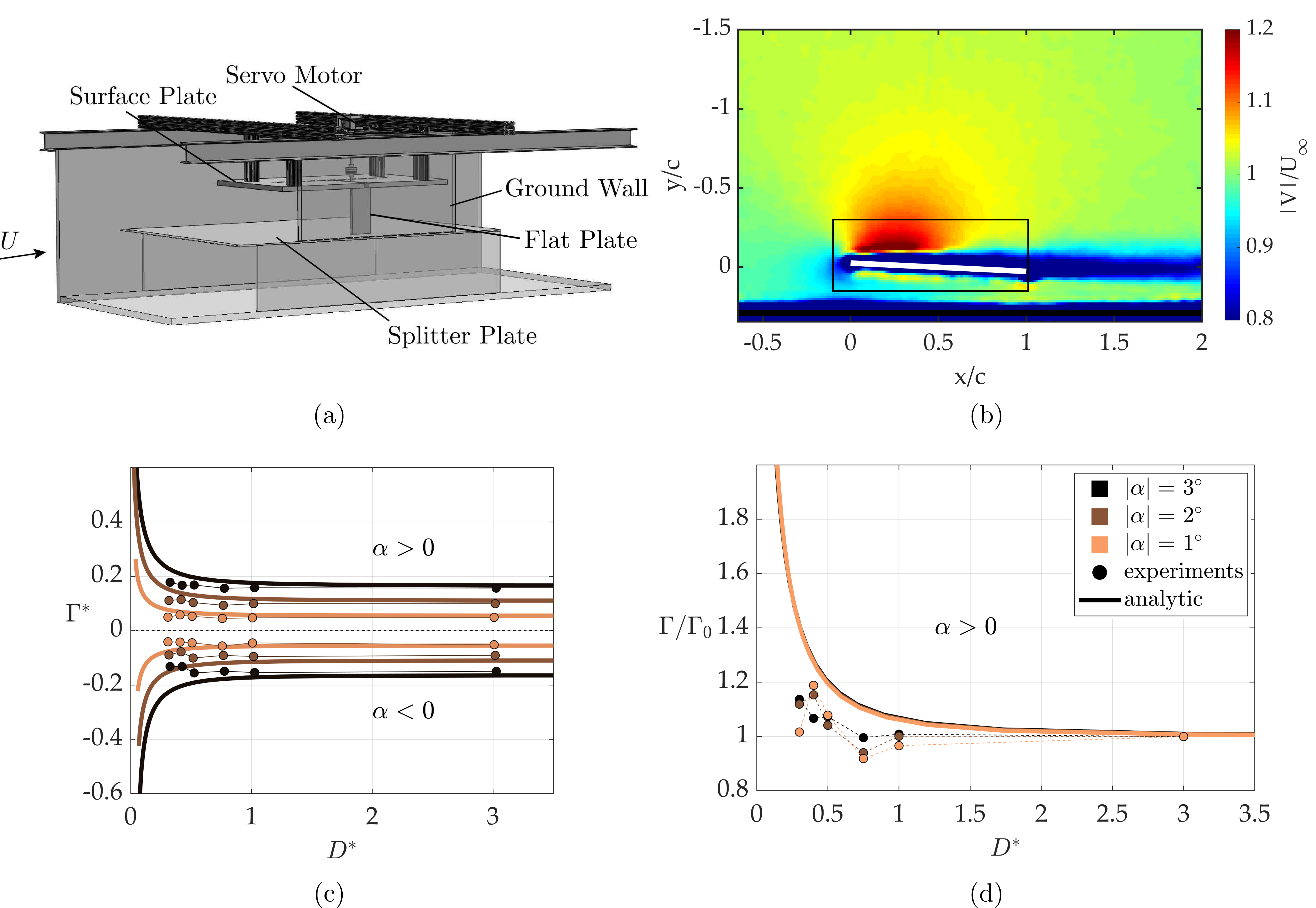}
\caption{A schematic of the experimental setup (a), a velocity-field around the flat plate located at $D^*=0.3$ and $\alpha = 3^{\circ}$ (b), a comparison of the dimensionless circulation (c), and normalized circulation with far-from-ground circulation (d) between experiments and the ideal solution. Square markers are color indicators for the corresponding $\alpha$.}
\label{fig:exp_part}
\end{figure}

The ground effect experiments were conducted in a closed-loop water channel with a test section of $4.9$ m long, $0.93$ m wide, and $0.61$ m deep. The flow was constrained from the bottom and the top with a splitter and surface plate to produce a nominally two-dimensional flow as shown in Figure \ref{fig:exp_part}a. Additionally, a vertical ground plane was installed on the side of the channel. The flat plate used in the experiments has a rectangular planform shape, a 4\% thick cross-section, and a chord and span length of $c = 0.095$ m and $s = 0.19$ m ($\AR = 2$), respectively. It was constructed from clear acrylic, and polished to be optically transparent for the flow-visualization. Throughout the experiments, the flow speed, $U$, was kept at $U=0.11$ m/s giving a chord-based Reynolds number of $Re = 11,800$. The flat plate was rotated with a servo-motor from its mid-chord to control the static angle of attack, $\alpha$. The ground distance $D^*=D/c$ was changed within the range of $0.3\leq D^* \leq 3$, and the angle of attack was varied within the range of $-3^{\circ} \leq \alpha \leq 3^{\circ}$ with $1^{\circ}$ increments for each ground distance.

Particle Image Velocimetry (PIV) data were acquired from the flow around the flat plate for each $(\alpha,D^*)$ case by using an Imager sCMOS camera ($2560 \times 2160$ pixels) and a 200 mJ/pulse Nd:YAG laser (EverGreen 200). The flow was seeded with 11 $\mu$m hollow metallic coated plastic sphere particles. At the beginning of each run, a digital signal was sent to the programmable timing unit, which triggers both camera and laser at the same time. Each flow-field was time-averaged over $100$ frames which were captured at $15$ Hz sampling rate. Four passes with two different window sizes were used in the vector calculations with a final interrogation window size of $48 \times 48$ pixels with 75\% overlap. The uncertainty in the instantaneous velocity fields is estimated to be between $1\%$ to $5\%$.%

The circulation for each $(\alpha,D^*)$ case was calculated from the velocity fields around a rectangular contour by using the line-integral method. The contour was chosen to include the flat plate but to avoid the boundary layer of the ground wall and shed wake from the plate as shown in Figure \ref{fig:exp_part}b. Non-zero circulation values calculated for $\alpha=0^{\circ}$ were subtracted from the circulation values obtained for the other angles of attack for each ground distance for direct comparison with the ideal solution. A comparison of the experiments is shown against the present model in Figure \ref{fig:exp_part}c and \ref{fig:exp_part}d. In Figure \ref{fig:exp_part}c, the dimensionless circulation ($\Gamma^* = \Gamma/(cU)$) is presented as a function of the dimensionless ground distance and angle of attack. For positive angles of attack, the dimensionless circulation is shown to increase as the ground distance decreases and as the angle of attack increases.  For the closest ground proximities and lowest positive angles of attack ($\alpha = 1^\circ,\, 2^\circ$) the dimensionless circulation decreases due to flow separation.  For the negative angles of attack, there is first a mild decrease in the dimensionless circulation as the ground distance decreases, in accordance with the ideal solution, and then there is an increase in circulation for the two closest ground distances due to flow separation.  For positive angles of attack the experiments show good agreement with $10-22\%$ deviation from the ideal solutions.  Figure \ref{fig:exp_part}d presents the circulation normalized with the far-from-ground circulation ($\Gamma_0$) for the positive angles of attack, where the model predictions for various angles of attack collapse onto nearly the same curve. Here, the corresponding experimental data follow the same trend as the ideal solutions with $8-12\%$ deviation, except for $D^*=0.3$ location with $21\%$ deviation. The deviation in the data from the predicted values is higher for the extreme-ground effect cases, $D^*\leq0.5$, which can be attributed to the viscous effects such as flow separation and interactions between the plate and ground boundary layers in the experiments, which are not accounted for in the ideal solution. 
\section{Conclusions} \label{Sec:conclusion}
We have presented a number of exact solutions for the flow past a wing in ground effect. The solutions are all expressed in terms of the $P$-function \eqref{Eq:SKprod}, which captures the doubly connected topology of the problem. Our solutions extend previous work by permitting more realistic geometries and flow phenomena.

We have also compared our exact, potential flow solution to experimental data in order to identify the missing physics in our model. The data indicate that boundary layer separation is significant, especially in the case of extreme ground effect. These effects are not currently accounted for by our model, although there are several ways that they could be included. The classical remedy is to use asymptotic analysis to solve for the flow in the boundary layers, and then match this to our potential flow solution \citep{vanDyke1964}. Alternatively, a modern approach is to couple our reduced-order potential flow model with a high-fidelity solution via data assimilation \citep{Darakananda2018a} to parsimoniously account for boundary layer separation and viscous effects. These extensions constitute future work.

\section{Acknowledgements}
P. J. B. and L. J. A. acknowledge support from EPSRC grants 1625902 and EP/P015980/1 respectively. P. J. B. would like to acknowledge useful conversations with Prof. Darren Crowdy and Dr. Rhodri Nelson at Imperial College London. For the experimental validation part, M. K. and K. W. M. would like to acknowledge the support by the Office of Naval Research under Program Director Dr. Robert Brizzolara on MURI grant number N00014-08-1-0642, as well as by the National Science Foundation under Program Director Dr. Ronald Joslin in Fluid Dynamics within CBET on NSF CAREER award number 1653181.

Declaration of Interests: The authors report no conflict of interest.

\bibliographystyle{jfm}
\bibliography{jfm-ref}

\begin{thebibliography}{4}
\expandafter\ifx\csname natexlab\endcsname\relax\def\natexlab#1{#1}\fi
\def\au#1{#1} \def\ed#1{#1} \def\yr#1{#1}\def\at#1{#1}\def\jt#1{\textit{#1}}
  \def\bt#1{#1}\def\bvol#1{\textbf{#1}} \def\vol#1{#1} \def\pg#1{#1}
  \def\publ#1{#1}\def\arxiv#1{#1}\def\org#1{#1}\def\st#1{\textit{#1}}

\bibitem[Crowdy(2008)]{Crowdy2008a}
{\sc \au{Crowdy, D.~G.}} \yr{2008}  \at{{Explicit solution for the potential
  flow due to an assembly of stirrers in an inviscid fluid}}.  \jt{J. Eng.
  Math.}  \bvol{62}~(4),  \pg{333--344}.

\bibitem[Crowdy(2012)]{Crowdy2012}
{\sc \au{Crowdy, D.~G.}} \yr{2012}  \at{{Conformal slit maps in applied
  mathematics}}.  \jt{ANZIAM J.}  \bvol{53}~(3),  \pg{171--189}.

\bibitem[Crowdy {\em et~al.\/}(2016)Crowdy, Kropf, Green \&
  Nasser]{Crowdy2016a}
{\sc \au{Crowdy, D.~G.}, \au{Kropf, E.~H.}, \au{Green, C.~C.} \& \au{Nasser, M.
  M.~S.}} \yr{2016}  \at{{The Schottky-Klein prime function: A theoretical and
  computational tool for applications}}.  \jt{IMA J. Appl. Math.}
  \bvol{81}~(3),  \pg{589--628}.

\bibitem[Crowdy \& Marshall(2006)]{Crowdy2006}
{\sc \au{Crowdy, D.~G.} \& \au{Marshall, J.}} \yr{2006}  \at{{Conformal
  mappings between canonical multiply connected domains}}.  \jt{Comput. Methods
  Funct. Theory}  \bvol{6}~(1),  \pg{59--76}.

\end{thebibliography}


\begin{thebibliography}{20}
\expandafter\ifx\csname natexlab\endcsname\relax\def\natexlab#1{#1}\fi
\def\au#1{#1} \def\ed#1{#1} \def\yr#1{#1}\def\at#1{#1}\def\jt#1{\textit{#1}}
  \def\bt#1{#1}\def\bvol#1{\textbf{#1}} \def\vol#1{#1} \def\pg#1{#1}
  \def\publ#1{#1}\def\arxiv#1{#1}\def\org#1{#1}\def\st#1{\textit{#1}}

\bibitem[Ahmed \& Goonaratne(2002)]{Ahmed2002}
{\sc \au{Ahmed, N.~A.} \& \au{Goonaratne, J.}} \yr{2002}  \at{{Lift
  augmentation of a low-aspect-ratio thick wing in ground effect}}.  \jt{J.
  Aircr.}  \bvol{39}~(2),  \pg{381--384}.

\bibitem[Baker(1897)]{Baker1897}
{\sc \au{Baker, H.~F.}} \yr{1897} {\em {Abelian functions: Abel's theorem and
  the allied theory of theta functions}\/}.  \publ{Cambridge University Press}.

\bibitem[Crowdy(2008)]{Crowdy2008}
{\sc \au{Crowdy, D.~G.}} \yr{2008}  \at{{The Schwarz problem in multiply
  connected domains and the Schottky–Klein prime function}}.  \jt{Complex
  Var. Elliptic Equations}  \bvol{53}~(3),  \pg{221--236}.

\bibitem[Crowdy(2010)]{Crowdy2010}
{\sc \au{Crowdy, D.~G}} \yr{2010}  \at{{A new calculus for two-dimensional
  vortex dynamics}}.  \jt{Theor. Comput. Fluid Dyn.}  \bvol{24}~(1-4),
  \pg{9--24}.

\bibitem[Crowdy(2020)]{Crowdy2020}
{\sc \au{Crowdy, D.~G.}} \yr{2020} {\em {Solving problems in multiply connected
  domains}\/}.  \publ{SIAM CBMS-NSF Regional Conference Series in Applied
  Mathematics}.

\bibitem[Crowdy \& Marshall(2006)]{Crowdy2006}
{\sc \au{Crowdy, D.~G.} \& \au{Marshall, J.}} \yr{2006}  \at{{Conformal
  mappings between canonical multiply connected domains}}.  \jt{Comput. Methods
  Funct. Theory}  \bvol{6}~(1),  \pg{59--76}.

\bibitem[Darakananda {\em et~al.\/}(2018)Darakananda, da~Silva, Colonius \&
  Eldredge]{Darakananda2018a}
{\sc \au{Darakananda, D.}, \au{da~Silva, A. F. de~C.}, \au{Colonius, T.} \&
  \au{Eldredge, J.~D.}} \yr{2018}  \at{{Data-assimilated low-order vortex
  modeling of separated flows}}.  \jt{Phys. Rev. Fluids}  \bvol{3}~(12),
  \pg{124701}.

\bibitem[Eldredge(2019)]{Eldredge2019a}
{\sc \au{Eldredge, J.~D.}} \yr{2019} {\em {Mathematical modeling of unsteady
  inviscid flows}\/},  \st{Interdisciplinary Applied Mathematics},
  \vol{vol.~50}.  \publ{Springer International Publishing}.

\bibitem[Iosilevskii(2008)]{iosilevskii2008asymptotic}
{\sc \au{Iosilevskii, G.}} \yr{2008}  \at{Asymptotic theory of an oscillating
  wing section in weak ground effect}.  \jt{European Journal of
  Mechanics-B/Fluids}  \bvol{27}~(4),  \pg{477--490}.

\bibitem[Joukowski(1910)]{Joukowsky1910}
{\sc \au{Joukowski, N.}} \yr{1910}  \at{{{\"{U}}ber die Konturen der
  Tragfl{\"{a}}chen der Drachenflieger}}.  \jt{Zeitschrift f{\"{u}}r
  Flugtechnik und Mot.}  \bvol{1},  \pg{281----284}.

\bibitem[Katz \& Plotkin(2001)]{Katz2001}
{\sc \au{Katz, J.} \& \au{Plotkin, A.}} \yr{2001} {\em {Low-Speed
  Aerodynamics}\/}.  \publ{Cambridge: Cambridge University Press},
  \arxiv{arXiv: arXiv:1011.1669v3}.

\bibitem[Kurt {\em et~al.\/}(2019)Kurt, Cochran-Carney, Zhong, Mivehchi, Quinn
  \& Moored]{Kurt2019}
{\sc \au{Kurt, M.}, \au{Cochran-Carney, J.}, \au{Zhong, Q.}, \au{Mivehchi, A.},
  \au{Quinn, D.~B.} \& \au{Moored, K.~W.}} \yr{2019}  \at{{Swimming freely near
  the ground leads to flow-mediated equilibrium altitudes}}.  \jt{J. Fluid
  Mech.}  \bvol{875},  \pg{R1}.

\bibitem[Michelin \& {Llewellyn Smith}(2009)]{Michelin2009}
{\sc \au{Michelin, S.} \& \au{{Llewellyn Smith}, S.~G.}} \yr{2009}  \at{{An
  unsteady point vortex method for coupled fluid-solid problems}}.  \jt{Theor.
  Comput. Fluid Dyn.}  \bvol{23}~(2),  \pg{127--153}.

\bibitem[Quinn {\em et~al.\/}(2014)Quinn, Moored, Dewey \&
  Smits]{quinn2014unsteady}
{\sc \au{Quinn, D.~B.}, \au{Moored, K.~W.}, \au{Dewey, P.~A.} \& \au{Smits,
  A.~J.}} \yr{2014}  \at{Unsteady propulsion near a solid boundary.}
  \jt{Journal of Fluid Mechanics}  \bvol{742},  \pg{152--170}.

\bibitem[Rayner(1991)]{rayner1991aerodynamics}
{\sc \au{Rayner, Jeremy M.~V.}} \yr{1991}  \at{On the aerodynamics of animal
  flight in ground effect}.  \jt{Philosophical Transactions of the Royal
  Society of London. Series B: Biological Sciences}  \bvol{334}~(1269),
  \pg{119--128}.

\bibitem[Rozhdestvensky(2006)]{rozhdestvensky2006wing}
{\sc \au{Rozhdestvensky, K.~V.}} \yr{2006}  \at{Wing-in-ground effect
  vehicles}.  \jt{Progress in Aerospace Sciences}  \bvol{42}~(3),
  \pg{211--283}.

\bibitem[Tomotika {\em et~al.\/}(1933)Tomotika, Nagamiya \&
  Takenouti]{Tomotika1933}
{\sc \au{Tomotika, S.}, \au{Nagamiya, T.} \& \au{Takenouti, Y.}} \yr{1933}
  \at{{The lift on a flat plate placed near a plane wall, with special
  reference to the effect of the ground upon the lift of a monoplane
  aerofoil}}.  \jt{Res. Inst. Tokyo} .

\bibitem[Tuck(1980)]{tuck1980nonlinear}
{\sc \au{Tuck, E.O.}} \yr{1980}  \at{A nonlinear unsteady one-dimensional
  theory for wings in extreme ground effect}.  \jt{Journal of Fluid Mechanics}
  \bvol{98}~(1),  \pg{33--47}.

\bibitem[{Van Dyke}(1964)]{vanDyke1964}
{\sc \au{{Van Dyke}, M.}} \yr{1964} {\em {Perturbation Methods in Fluid
  Mechanics}\/}.  \publ{New York: Academic press}.

\bibitem[Widnall \& Barrows(1970)]{Widnall1970}
{\sc \au{Widnall, S.~E.} \& \au{Barrows, T.~M.}} \yr{1970}  \at{{An analytic
  solution for two-and three-dimensional wings in ground effect}}.  \jt{J .
  Fluid Mech}  \bvol{41}~(4),  \pg{769--792}.

\end{thebibliography}

\tikzexternaldisable

\setcounter{section}{0}
\setcounter{subsection}{0}
\setcounter{equation}{0}

\renewcommand\thesection{S:\arabic{section}}      
\renewcommand\thefigure{\thesection.\arabic{figure}}    
\renewcommand\theequation{\thesection.\arabic{equation}}
\newpage

\setcounter{page}{1}

\shorttitle{Exact solutions for ground effect -- supplementary material}
\vspace*{14pt}
\begin{center}
{
\LARGE\bfseries
{Exact solutions for ground effect -- supplementary material}
  {\large\bfseries\baselineskip=18pt
  \vskip 14pt
     {Peter. J. Baddoo, Melike. Kurt, Lorna. J. Ayton \& Keith W. Moored}\par}%
}
\end{center}
\vspace*{30pt}

This supplementary material includes several details of calculations and methods used in the main paper.

\section{The $P$-function} \label{Sec:Pfun}

The $P$-function was introduced in section \ref{Sec:solution}. Therein, we presented an infinite product representation. Whilst the product form offers physical intuition, it also has poor convergence properties. For numerical computations it is more convenient to express the $P$-function as the rapidly convergent Laurent series \citepSupp{Crowdy2012}
	\begin{align}
	P(\zeta)  &= C \sum_{n=-\infty}^\infty (-1)^nq^{n(n-1)}\zeta^n, \qquad\qquad C = \frac{\prod_{n=1}^\infty(1+q^{2n})^2}{\sum_{n=1}^\infty q^{n(n-1)}}. \label{Eq:SKsum}
	\end{align}
This series converges quickly except when $q$ is very close to unity. In that case, we recommend calculating the $P$-function by using the spectral method in \citeSupp{Crowdy2016a}, for which software is freely available (\url{https://github.com/ACCA-Imperial/SKPrime}).

It is possible to show from the product formula \ref{Eq:SKprod} that the $P$-function has the following properties
\begin{align}
\addtocounter{equation}{0}
    P(\zeta^{-1})&= -\zeta^{-1} P(\zeta),\label{Eq:pRel2} \tag{\theequation.a}\\
        P(q^2\zeta) &=-\zeta^{-1} P(\zeta), \label{Eq:pRel1} \tag{\theequation.b}\\
    \frac{P(q^2 \gamma_1^{-1} \zeta)}{P(q^2  \gamma_2^{-1}\zeta)} &= \frac{\gamma_1}{\gamma_2} \cdot \frac{P( \gamma_1^{-1}\zeta)}{P( \gamma_2^{-1}\zeta)}.\label{Eq:pRel3} \tag{\theequation.c}
\end{align}
These properties will be used in the derivation of the ground effect conformal maps.
\section{Derivations of conformal mappings} \label{Sec:confProof}

In this section we derive the conformal mappings presented in section \ref{Sec:conformal}.

\subsubsection{Derivation of circular wing map}
\begin{figure}
\centering
\niceFigure{\mywidth}{\myheight}{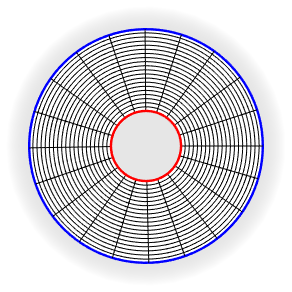}{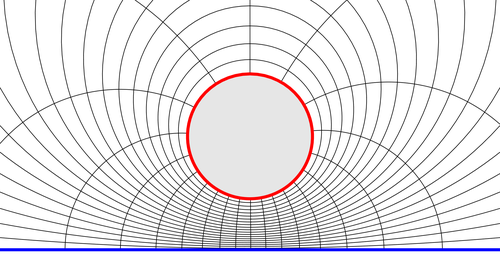}
	\caption{An example of a circular wing map \eqref{Eq:Mobius}.}
	\label{Fig:mob}
\end{figure}

Given that \eqref{Eq:Mobius} is a M\"obius mapping, it maps circlines to circlines. Since $\zeta=1$ is a simple pole of the mapping, the boundary circle $C_0$ is mapped to a circle of infinite radius, i.e. a line. It is straight foward to show that this line is the real axis. Additionally, $C_1$ must be mapped to a circle, and it can be shown that the image of $C_1$ is centered at $\i (q +q^{-1})/2$ and has unit radius. A typical circular wing map is illustrated in figure \ref{Fig:mob}
\subsubsection{Derivation of flat plate map}
In order to derive the flat plate map \eqref{Eq:ConfMap}, we will show that $f$ takes constant phase on $C_0$ and $C_1$ when the shifting constant $s$ vanishes.

We write $A = \tilde{A} \e^{-\i \alpha}$ where $A$ is pure real. The complex conjugate of $f$ is then given by
\begin{align}
    \overline{f(\zeta)} &= \tilde{A} \e^{\i \alpha} \frac{{P}(\e^{-2 \i \alpha}\overline{\zeta} )}{P(\overline{\zeta})}. \label{Eq:fbar1}
\end{align}
For $\zeta \in C_0$, we may write $\bar{\zeta}=1/\zeta$ and use \eqref{Eq:pRel2} to transform \eqref{Eq:fbar1} into 
\begin{align}
    \overline{f(\zeta)} &= \tilde{A} \e^{-\i \alpha} \frac{{P}(\e^{2 \i \alpha}{\zeta})}{P({\zeta})} = f(\zeta).
\end{align}
Therefore, $f$ is pure real for $\zeta \in C_0$. Since $f$ contains a simple pole and is univalent, $f(\zeta)$ spans the entire real line.

We now consider the case where $\zeta \in C_1$. In this case, we have $\bar{\zeta} = q^2/\zeta$. Combining this fact with \eqref{Eq:pRel3} in \eqref{Eq:fbar1} yields
\begin{align}
    \overline{f(\zeta)} & = \tilde{A} \e^{\i \alpha} \frac{P(\e^{-2 \i \alpha}/{\zeta})}{P(1/\zeta)}.
\end{align}
A further application of \eqref{Eq:pRel2} yields $\overline{f} = \e^{2\i \alpha}f$. Therefore, $\arg[f(\zeta)]=-\alpha$ for $\zeta \in C_1$, so $f$ maps $C_1$ to a slit inclined at an angle of $-\alpha$ to the real axis, which corresponds to a flat plate at angle of attack $\alpha$.

Typical flat plate wing maps are illustrated in figures \ref{Fig:flat} and \ref{Fig:flat0}.
{
\begin{figure}
\centering
\niceFigure{\mywidth}{\myheight}{figures/zet-map1.png}{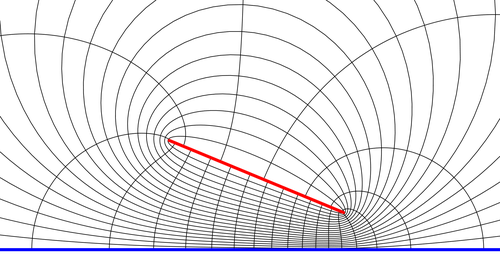}
	\caption{An example of a flat plate wing mapping \eqref{Eq:ConfMap}.}
	\label{Fig:flat}
\end{figure}
	
\begin{figure}
\centering
\niceFigure{\mywidth}{\myheight}{figures/zet-map1.png}{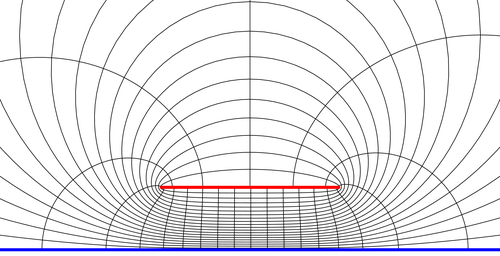}
	\caption{An example of a flat plate wing mapping at zero angle of attack \eqref{Eq:confMap0}.}
	\label{Fig:flat0}
\end{figure}
}
\subsubsection{Derivation of circular arc wing map}
We first note that the annulus may be mapped to a circle with a circular arc slit by the mapping \citepSupp{Crowdy2006}
\begin{align}
    \eta(\zeta,\gamma) &= \frac{P(\zeta/\gamma)}{|\gamma|P(\zeta\overline{\gamma})}.
\end{align}
for any $\gamma \in D_\zeta$. This mapping transplants the unit disc onto itself, and $C_1$ onto a circular arc slit. We now take another M\"obius mapping to map the unit circle to the real axis and the circular slit to another circular slit on the upper half plane:
\begin{align*}
f(\zeta) = \frac{B}{\eta(1,\gamma) - \eta(\zeta,\gamma)},    
\end{align*}
where $B$ is a constant to rotate and scale the map. Note that we have ensured that there is a simple pole at $\zeta=1$. Composing the mappings gives
\begin{align*}
    f(\zeta) &= \frac{A P(\zeta\overline{\gamma})}{P(\zeta/\gamma)P(\overline{\gamma}) - P(\zeta/ \gamma)P(\zeta \overline{\gamma})},
\end{align*}
for another constant $A$.

\begin{figure}
\centering
\niceFigure{\mywidth}{\myheight}{figures/zet-map1.png}{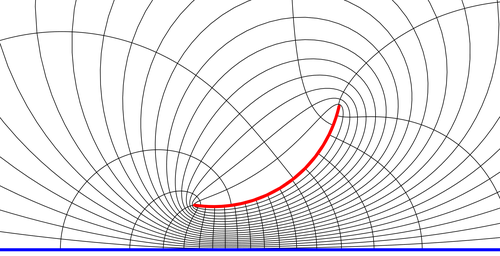}
	\caption{An example of a circular arc wing map \eqref{Eq:circArcMap}.}
	\label{Fig:circArc}
\end{figure}

A typical circular arc wing map is illustrated in figure \ref{Fig:circArc}.
\subsubsection{Derivation of centered circular arc wing map}
In order to derive the centered circular arc wing map, we first note that we may write
\begin{align*}
    P(\sqrt{ \zeta})P(-\sqrt{ \zeta}) & = P_2( \zeta),
\end{align*}
where
\begin{align}
    P_2(\zeta) =  (1 - \zeta) \prod_{k=1}^{\infty} \left(1 - q^{4k} \zeta \right) \cdot \left( 1- q^{4k} \zeta^{-1}\right). \label{Eq:SK2prod}
\end{align}
The interpretation of $P_2$ is that it is the $P$-function for an annulus where the ratio of exterior to interior radius is $q^2$. The analogous forms of (\ref{Eq:pRel2}, \ref{Eq:pRel1}, \ref{Eq:pRel3}) are
\begin{align}
\addtocounter{equation}{0}
    P_2(\zeta^{-1})&= -\zeta^{-1} P_2(\zeta), \label{Eq:p2Rel2}\tag{\theequation.a} \\
        P_2(q^4\zeta) &=-\zeta^{-1} P_2(\zeta),\label{Eq:p2Rel1}  \tag{\theequation.b}\\
    \frac{P_2(q^4 \gamma_1^{-1} \zeta)}{P_2(q^4  \gamma_2^{-1}\zeta)} &= \frac{\gamma_1}{\gamma_2} \cdot \frac{P_2( \gamma_1^{-1}\zeta)}{P_2( \gamma_2^{-1}\zeta)}.\label{Eq:p2Rel3}  \tag{\theequation.c}
\end{align}
The centered circular arc wing map \eqref{Eq:centCircMap} may therefore be written as
\begin{align}
    f(\zeta) & = \tilde{A} \e^{-\i \phi} \frac{P_2({\zeta}\e^{2\i \phi})P_2(q^2 {\zeta})}{P_2(q^2 {\zeta} \e^{2\i \phi} ) P_2({\zeta})}, \label{Eq:circSlit}
\end{align}
where $A = \tilde{A} \e^{- \i \phi}$. The first step in the derivation of the circular arc wing map is to see that \eqref{Eq:circSlit} is a circular slit map on $q<|\zeta|<1/q$: the circles $|\zeta|=q$ and $|\zeta|=1/q$ are transplanted to two circular arc slits. We first consider the circle $|\zeta| = q$ so that
\begin{align*}
    f(\zeta) = \tilde{A} \e^{-\i \phi} \frac{P_2(q^4 /\overline{\zeta})}{P_2(q^4\e^{2\i \phi}/\overline{\zeta} )} \cdot \frac{P_2(q^2\e^{2\i \phi}/\overline{\zeta})}{P_2(q^2/\overline{\zeta})},
\end{align*}
since $\overline{\zeta} = q^2/\zeta$. Application of \eqref{Eq:p2Rel3} yields
\begin{align*}
    f(\zeta) = \tilde{A} \e^{-\i \phi} \frac{P_2(1 /\overline{\zeta})}{P_2(\e^{2\i \phi}/\overline{\zeta} )} \cdot \frac{P_2(\e^{2\i \phi}/(q^2\overline{\zeta}))}{P_2(1/(q^2\overline{\zeta}))}.
\end{align*}
By taking the complex conjugate and applying \eqref{Eq:p2Rel2}, we see that 
\begin{align*}
f(\zeta) \overline{f(\zeta)} = |f(\zeta)|^2= 1.
\end{align*}
Therefore the image of $|\zeta| = q$ is a circular arc of unit radius. A similar procedure may be applied to show that the circle $|\zeta|= 1/q$ is also a circular arc.

We now show that the map is anti-symmetric when the argument is reflected in the unit circle. Reflecting the argument of $f$ yields
\begin{align*}
    f(1/\overline{\zeta}) &= \tilde{A} \e^{-\i \phi} \frac{P_2({\e^{2\i \phi}/\overline{\zeta}})}{P_2(1/\overline{\zeta})} \cdot  \frac{P_2(q^2/\overline{\zeta})}{P_2(q^2 \e^{2\i \phi}/\overline{\zeta})}.
\end{align*}
Applying \eqref{Eq:p2Rel3} to the second fraction gives
\begin{align*}
    f(1/\overline{\zeta}) &= \tilde{A} \e^{\i \phi} \frac{P_2({\e^{2\i \phi}/\overline{\zeta}})}{P_2(1/\overline{\zeta})}  \cdot  \frac{P_2(1/(q^2\overline{\zeta}))}{P(\e^{2\i \phi}/(q^2\overline{\zeta}))}.
\end{align*}
Now applying \eqref{Eq:p2Rel2} to each $P_2$ yields $f(\zeta) = \overline{f(1/\overline{\zeta})}$. In other words, reflecting the point $\zeta$ in the circle $|\zeta| = 1$ results in a reflection of $f$ in the real axis. Since $f$ contains a simple pole at $\zeta=1$, the unit disc $|\zeta|=1$ is mapped to the entire real axis.

\begin{figure}
\centering
\niceFigure{\mywidth}{\myheight}{figures/zet-map1.png}{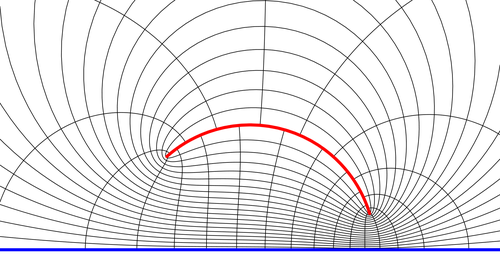}
	\caption{An example of a centered circular arc wing map \eqref{Eq:centCircMap}.}
	\label{Fig:centCircArc}
\end{figure}

A typical centered circular arc wing map is illustrated in figure \ref{Fig:centCircArc}.

\section{Residues of conformal maps} \label{Sec:residue}
In this section we provide the residues of each conformal map at the simple pole $\zeta=1$. The residue of the circular wing map  \eqref{Eq:Mobius} is
\begin{align*}
    a_{\infty} = \frac{q^2-1}{\i q}.
\end{align*}
The residue of the flat plate map \eqref{Eq:ConfMap} at angle of attack $\alpha$ is
\begin{align*}
    a_{\infty} &= \frac{-A  P(\e^{2 \i \alpha}) }{L}.
\end{align*}
In the degenerate case where the angle of attack is zero \eqref{Eq:confMap0}, the residue is
\begin{align*}
    a_{\infty} &= \i A.
\end{align*}
The residue for the circular arc wing map \eqref{Eq:circArcMap} is 
\begin{align*}
    a_{\infty} &= \frac{-\gamma A P(\overline{\gamma})}{P^\prime(1/\gamma)P(\overline{\gamma}) - |\gamma|^2 P^\prime(\overline{\gamma})P(1/\gamma)}.
\end{align*}
The residue for the centered circular arc wing map \eqref{Eq:centCircMap} is
\begin{align*}
    a_{\infty} &= \frac{A}{L} \cdot \frac{ 2P(\e^{\i \phi})P(-\e^{\i \phi})P(q)P(-q)}{P(q \e^{\i \phi}) P(-q \e^{ \i \phi})P(-1)} .
\end{align*}

\section{Derivation of kinematic boundary condition for moving plate} \label{Sec:kinematic-sup}
In this section we derive the kinematic boundary condition in the case where the wing executes rigid body motions. We do so by adapting the analysis of \citeSupp{Crowdy2008a}. Since the ground is stationary, it must represent a streamline. Therefore, the complex potential takes constant imaginary part on $C_0$, which we may take to be zero without loss of generality. However, on the wing, the kinematic boundary condition states that fluid on the wing must move at the same velocity as the wing itself:
\begin{align}
\mathbf{u}\cdot \mathbf{n} &= \mathbf{U} \cdot \mathbf{n}, \label{Eq:KinBCVec}
\end{align}
where $\mathbf{u}$ represents the fluid velocity, $\mathbf{n}$ represents the outward normal direction, and $\mathbf{U}$ represents the velocity of wing. The normal vector $\mathbf{n}$ may be written as $-\i \d z / \d s$ where $s$ is the arc length, so that the kinematic boundary condition \eqref{Eq:KinBCVec} may be written as 
\begin{align*}
\Re\left[\bar{u}(s) \times -\i \frac{\d z}{\d s}\right] &= \Re \left[ \bar{U}(s)\times -\i \frac{\d z}{\d s}\right].
\end{align*}
In the physical plane, the complex potential due to the motion of the wing is given by $w_M$. The kinematic condition therefore becomes
\begin{align}
\Re\left[\frac{\d w_M}{\d z} \times -\i \frac{\d z}{\d s}\right] &= \Re \left[ \bar{U}(s)\times -\i \frac{\d z}{\d s}\right]. \label{Eq:kin1}
\end{align}
The first term may be simplified by the chain rule to
\begin{align*}
\Re\left[ -\i \frac{\d w_M}{\d s} \right] &= \Re \left[-\i \bar{U}(s) \frac{\d z}{\d s}\right].
\end{align*}
We now note that the wing may be parametrised as
\begin{align*}
z = D(t) + \eta(s)\e^{-\i \alpha(t)}.
\end{align*}
The velocity of the plate may therefore be written as
\begin{align*}
U(s) &= \dot{D}(t) - \i \dot{\alpha}(t) \eta(s) \e^{-\i \alpha(t)} =  \dot{D}(t) - \i \dot{\alpha}(t) \left(z-D(t)\right),
\end{align*}
Therefore, the kinematic condition \eqref{Eq:kin1} becomes
\begin{align}
\Re\left[ -\i \frac{\d w_M}{\d s} \right] &= \Re \left[-\i \dot{\bar{D}}(t)\frac{\d z}{\d s} + \dot{\alpha}(t) \left(\bar{z}-\bar{D}(t)\right) \frac{\d z}{\d s}\right]. \label{Eq:kin2}
\end{align}
Noting that 
\begin{align*}
\frac{\d }{\d s} \left| z-D\right|^2 &= 2 \Re\left[\frac{\d z}{\d s} \left(\bar{z}- \bar{D}\right)\right],
\end{align*}
we may now integrate the kinematic condition  \eqref{Eq:kin2} with respect to $s$ to obtain
\begin{align}
\Re\left[ -\i w_M \right] &= \Re \left[-\i \dot{\bar{D}}(z-D) - \frac{\dot{\alpha}}{2} \left| z-D\right|^2 \right] + I, \label{Eq:IntKinBC}
\end{align}
for a constant of integration $I$ which will be chosen to comply with a compatibility condition to be defined later.

We write $W_M(\zeta)=w_M(z(\zeta))$ and translate \eqref{Eq:IntKinBC} into the canonical circular domain to obtain the condition
\begin{align}
\Im\left[ W_M(\zeta) \right] &= \Im \left[ \dot{\bar{D}}(z(\zeta)-D) + \i\frac{\dot{\alpha}}{2} \left| z(\zeta)-D\right|^2 \right] + I \equiv M\left(\zeta,\bar{\zeta} \right). \qquad \textnormal{for } \zeta \in C_1. \label{Eq:IntKinBCZetaImag}
\end{align}
Finally, the constant $I$ is given by enforcing a compatibility condition:
\begin{align}
    I =-\frac{1}{2 \pi \i} \oint_{C_1} \Im \left[ \dot{\bar{D}}(z(\zeta^\prime)-D) + \i \frac{\dot{\alpha}}{2} \left| z(\zeta^\prime)-D\right|^2 \right] \frac{\d \zeta^\prime}{\zeta^\prime}. \label{Eq:Idef}
\end{align}

\bibliographystyleSupp{jfm}
\bibliographySupp{jfm-ref}

\end{document}